# Solutions of the nonrelativistic wave equation with position-dependent effective mass

A. D. Alhaidari

*Physics Department, King Fahd University of Petroleum & Minerals, Box 5047, Dhahran 31261, Saudi Arabia*
E-mail: haidari@mailaps.org

Given a spatially dependent mass distribution we obtain potential functions for exactly solvable nonrelativistic problems. The energy spectrum of the bound states and their wavefunctions are written down explicitly. This is accomplished by mapping the wave equation for these systems into well-known exactly solvable Schrödinger equations with constant mass using point canonical transformation. The Oscillator, Coulomb, and Morse class of potentials are considered.

## I. INTRODUCTION

Exact solutions of the wave equation are important because of the conceptual understanding of physics that can only be brought about by such solutions. These solutions are also valuable means for checking and improving models and numerical methods being introduced for solving complicated physical problems. Exactly solvable problems fall within distinct classes of shape invariant potentials [1]. Each class carries a representation of a dynamical symmetry group. All potentials in a given class, along with their corresponding solutions (energy spectrum and wavefunctions), can be mapped into one another by point canonical transformation (PCT) [2]. Henceforth, only one problem (the "reference problem") in a given class needs to be solved to obtain solutions of all others in the class. PCT maintains the functional form of the problem (i.e. shape invariance of the potential). In other words, it leaves the form of the wave equation invariant. As a result, a correspondence map among the potential parameters, angular momentum, and energy of the two problems (the new and reference problem) is obtained. Using the parameter substitution map and the bound states spectrum of the reference problem one can easily and directly obtain the spectra of all other potentials in the class. Moreover, the wavefunctions are obtained by simple transformations of those of the reference problem. An alternative approach is to start with a problem whose exact solution is well established then apply to it PCTs that preserve the structure of the wave equation resulting in exactly solvable problems that belong to the same class as that of the original (reference) problem. Thus the reference problem acts like a seed for generating new exact solutions. This approach is suitable in the search for solutions of a given wave equation by making an exhaustive study of all PCTs that maintain shape invariance of the given wave equation.

Supersymmetric quantum mechanics [3] and potential algebras [4] are two among other methods beside PCT which are used in the search for exact solutions of the wave equation. In nonrelativistic quantum mechanics with *constant mass*, this development was carried out over the years by many authors where several classes of shape invariant potentials being accounted for and tabulated [1]. It was also extended to new classes of conditionally exactly [5] and quasi exactly [6] solvable potentials where all or, respectively, part of the energy spectrum is known. The relativistic extension of this development has been established only recently [7-9]. One of these relativistic classes, which includes the Dirac-Oscillator, Dirac-Coulomb and Dirac-Morse potentials, was investigated and found to carry a representation of a superalgebra which is a graded extension of so(2,1) Lie algebra. Each of these relativistic problems could be mapped into one another by an "extended point canonical transformation (XPCT)" [9].

Recently, several contributions have emerged in the literature where some of the above-mentioned developments in nonrelativistic quantum mechanics were extended to the case of spatially dependent mass distribution [10-14]. The motivation for obtaining exact solutions of the wave equation with position dependent mass comes from the wide range of applications of these solutions in various areas of material science and condensed matter. Such applications are found in the study of electronic properties of semiconductors [15], quantum dots [16], $^{3}$He clusters [17], quantum liquids [18] and semiconductor heterostructures [19], ...etc. The one dimensional Schrödinger equation with smooth mass and potential steps was solved exactly by L. Dekar *et al* [10]. The usual formalism of supersymmetric quantum mechanics was extended by A. R. Plastino *et al* to the Schrödinger equation with position dependent effective mass [11]. Shape invariance was also addressed in this setting and the energy spectra were obtained algebraically for several examples. Coordinate transformation in supersymmetric quantum mechanics were used in [12] to generate isospectral potentials with position dependent mass. The ordering ambiguity of the mass and momentum operator and its effect on the exact solutions was addressed in [13] where several examples are considered. so(2,1) Lie algebra as a spectrum generating algebra and as a potential algebra was used in [14] to obtain exact solutions of the effective mass wave equation.

In all cited work above, except for reference [10], the main concern was primarily focused on obtaining the energy spectrum and potential function for a given mass distribution. The wavefunctions, on the other hand, were either written down formally as solutions to a Schrödinger equation with constant mass whose potential is quite involved, or few lower states were obtained by successive application of the raising operators on the ground state which is often not too difficult to obtain. The exception is in the work by L. Dekar *et al* [10] where the authors went admirably at great length to solve for the wavefunction in terms of the Heun's function after transforming the wave equation to a Fuchsian equation with four singularities. Furthermore, in most of the developments mentioned above no integrability condition on the effective mass or potential functions was given. That is, no constraints were imposed on the mass or the potential such that the problem becomes exactly solvable. Nor a general criterion was given for the proper selection of these functions such that they become compatible with the formalism. That's why most of the examples given there are either limited or repetitive. Here we follow an alternative approach to the solution of the problem: Starting from a well known exactly solvable Schrödinger equation with constant mass (the reference problem; e.g., the harmonic oscillator) we apply a point canonical transformation to map it into a wave equation with the given spatially dependent mass ("the target problem"). Therefore, the resulting map will give not only the energy spectrum of the target problem but also the corresponding wavefunctions in terms of those of the reference problem. Obviously, these will be written in terms of the familiar orthogonal polynomials (e.g., Laguerre, Jacobi, etc.) but with arguments whose expressions may not always be simple but easily derivable. The canonical map will also give the associated target potential function belonging to the class which is defined by the reference potential and the target mass function. Moreover, we find a constraint for the analytic solvability of the problem which is that the square root of the mass function should be analytically integrable. We will consider three classes of reference potentials: the Oscillator, Coulomb and Morse. The details of the method will be presented in the case of the oscillator class while only a brief summary will be given for the other two classes. It is to be noted that the same development can be applied to other classes of shape invariant potentials. The class that includes Rosen-Morse, Scarf, Eckart and Pöschl-Teller potentials could be of prime significance.

The paper is organized as follows. In section II, the general development of the formalism will be presented and applied to the oscillator class. A summary in the case of the Coulomb and Morse classes will be given in section III where we also address the problem in three dimensions. To demonstrate the use of the method and illustrate the utility of our findings, few examples will be given in the Appendix for a selected set of mass distributions satisfying, of course, the solvability constraint.

## II. ACTION OF THE PCT MAP ON THE SCHRÖDINGER EQUATION APPLICATION TO THE OSCILLATOR CLASS

Following the consensus among the majority of the work done on the subject we will adopt the symmetric ordering of the momentum and mass as given by the following Hamiltonian:

$$H = \frac{1}{2}\left[\vec{P}\frac{1}{M(\vec{r})}\vec{P}\right] + V(\vec{r}) = -\frac{\hbar^2}{2m_0}\left[\vec{\nabla}\frac{1}{m(\vec{r})}\vec{\nabla}\right] + V(\vec{r}) \qquad (2.1)$$

where $m(\vec{r})$ and $V(\vec{r})$ are real functions of the configuration space coordinates. Using atomic units ($m_0 = \hbar = 1$), this Hamiltonian will produce the following time independent wave equation in one dimension

$$\left\{\frac{d^2}{dx^2} - \frac{m'}{m}\frac{d}{dx} - 2m[V(x) - E]\right\}\phi(x) = 0 \qquad (2.2)$$

where $E$ is the energy eigenvalue and $m' \equiv dm/dx$. On the other hand, the one-dimensional time independent Schrödinger wave equation with constant mass, potential function $\mathcal{V}$, and energy $\mathcal{E}$ reads

$$\left\{\frac{d^2}{dy^2} - 2[\mathcal{V}(y) - \mathcal{E}]\right\}\psi(y) = 0 \qquad (2.3)$$

We apply to this last equation the following transformation:
$$y = q(x), \quad \psi(y) = g(x)\phi(x) \qquad (2.4)$$

If the result is a mapping into Eq. (2.2) then this transformation will be referred to as "point canonical transformation (PCT)". Now, the action of (2.4) on Eq. (2.3) gives the following 2nd order differential equation

$$\left\{\frac{d^2}{dx^2} + \left(2\frac{g'}{g} - \frac{q''}{q'}\right)\frac{d}{dx} + \left(\frac{g''}{g} - \frac{q''}{q'}\frac{g'}{g}\right) - 2(q')^2[\mathcal{V}(q(x)) - \mathcal{E}]\right\}\phi(x) = 0 \qquad (2.5)$$

By identifying this with Eq. (2.2) we obtain the following conditions on the transformation (2.4) to be a PCT:

$$g(x) = \sqrt{q'/m} \tag{2.6}$$

$$V(x) - E = \frac{(q')^2}{m}[\mathcal{V}(q) - \mathcal{E}] + \frac{1}{m}F(m,q) \tag{2.7}$$

where

$$F(m,q) = \frac{1}{4}\left[\frac{m''}{m} - \frac{3}{2}\left(\frac{m'}{m}\right)^2 - \frac{q'''}{q'} + \frac{3}{2}\left(\frac{q''}{q'}\right)^2\right] \tag{2.8}$$

We define the following dimensionless integral (modulo a constant), which will appear frequently in subsequent developments:

$$\mu(x) = (1/\tau)\int\sqrt{m(x)}dx \tag{2.9}$$

where $\tau$ is a length scale parameter. Now, given $m(x)$ then a choice of transformation function $q(x)$ will determine $g(x)$, as given by Eq. (2.6), and will be used in Eq. (2.7) to deduce the energy spectrum $E$ and potential function $V(x)$ for the target problem. To this end, we consider two possibilities that will result in a constant term on the right hand side of Eq. (2.7), which will be identified with the energy $E$. The first is by taking $(q')^2 = m$ giving $q(x) = \tau\mu(x)$ and resulting in the following energy spectrum, potential and wavefunction for the target problem

$$E_n = \mathcal{E}_n$$

$$V(x) = \mathcal{V}(\tau\mu(x)) + \frac{1}{8m}\left[\frac{m''}{m} - \frac{7}{4}\left(\frac{m'}{m}\right)^2\right] \tag{2.10}$$

$$\phi_n(x) = m(x)^{1/4}\psi_n(\tau\mu(x))$$

The second possibility is by taking $(q')^2\mathcal{V}(q) = \pm m/\sigma^2$ giving $q(x) = \mathcal{W}^{-1}(\tau\mu(x)/\sigma)$, where $\mathcal{W}(y) = \int\sqrt{\pm\mathcal{V}(y)}dy$ and $\sigma$ is another scale parameter. The results in this case are as follows:

$$E_n = -1/\sigma^2$$

$$V(x) = -\frac{\mathcal{E}_n/\sigma^2}{\mathcal{V}(q(x))} + \frac{1}{8m}\left[\frac{m''}{m} - \frac{7}{4}\left(\frac{m'}{m}\right)^2\right] + \frac{1}{8\sigma^2\mathcal{V}}\left[\frac{\tilde{\tilde{\mathcal{V}}}}{\mathcal{V}} - \frac{5}{4}\left(\frac{\tilde{\mathcal{V}}}{\mathcal{V}}\right)^2\right] \tag{2.11}$$

$$\phi_n(x) = \left[\sigma^2 m(x)\mathcal{V}(q(x))\right]^{1/4}\psi_n(q(x))$$

where $\tilde{\mathcal{V}}(q) \equiv d\mathcal{V}(q)/dq$. Requiring that the first term in the potential expression above be independent of the index $n$ will result in a constraint on the parameter $\sigma$ relating it to the reference energy spectrum $\mathcal{E}_n$. Due to the fact that the expression of $F(m,q)$ is homogeneous in $q$ with zero degree, as seen in Eq. (2.8), then the last term in the expression of the potential in Eq. (2.11) will always be independent of $\sigma$. Consequently, $\sigma$ will not appear in the target problem quantities, thus acting as a dummy parameter introduced only to facilitate the calculation.

We should point out, however, that it is generally possible that other choices of $q(x)$ might be found which could result in a constant term on the right hand side of Eq. (2.7) to be interpreted as the energy $E$ and, thus, resulting in other solutions.

The problem can, therefore, be stated as follows:

"Given a position dependent mass $m(x)$ and a well known exactly solvable system, described by Eq. (2.3), we seek to find all potentials, $V(x)$, in the class defined by $\mathcal{V}$ and $m$ that will result in an exact solution of the system described by Eq. (2.2). That is, we search for all PCT functions, $q(x)$, satisfying Eq. (2.7) and thus *each* resulting in a conical map which gives $V(x)$, $E_n$ and $\phi_n(x)$ for a corresponding target system".

We start by illustrating the solution of the problem for the Oscillator class. That is by taking $\mathcal{V}(y) = \tfrac{1}{2}\lambda^4 y^2$, where $\lambda$ is the oscillator strength parameter. The energy eigenvalues and the corresponding wavefunctions for the reference problem are [20]:

$$\begin{aligned}\mathcal{E}_n &= \lambda^2(n+\tfrac{1}{2}) \\ \psi_n(y) &= a_n e^{-\lambda^2 y^2/2} H_n(\lambda y)\end{aligned} \quad ; \quad n = 0,1,2,\ldots \qquad (2.12)$$

where $H_n(x)$ is the Hermite polynomial [21] and $a_n$ is the normalization constant. The two possible solutions are:

(Oscillator-1): $(q')^2 = m$

Substituting in Eq. (2.10) gives the following target potential

$$V(x) = \frac{1}{2}\alpha^2 \mu(x)^2 + \frac{1}{8m}\left[\frac{m''}{m} - \frac{7}{4}\left(\frac{m'}{m}\right)^2\right] \qquad (2.13)$$

where $\alpha$ is a real parameter. The energy spectrum and associated wavefunctions of the target problem are, respectively, as follows:

$$\begin{aligned}E_n &= (\alpha/\tau)(n+\tfrac{1}{2}) \\ \phi_n(x) &= A_n m(x)^{1/4} e^{-\alpha\tau\mu(x)^2/2} H_n\left(\sqrt{\alpha\tau}\,\mu(x)\right)\end{aligned} \quad ; \quad n = 0,1,2\ldots \qquad (2.14)$$

where $A_n$ is the normalization constant. The other solution is as follows.

(Oscillator-2): $(q')^2 \mathcal{V}(q) = m/\sigma^2$

For the oscillator class $\mathcal{W}(y) = \left(\lambda^2/2\sqrt{2}\right) y^2$ giving $q(x) = \left(2^{3/4}\lambda^{-1}\sqrt{\tau/\sigma}\right)\sqrt{\mu(x)}$. Substituting in Eq. (2.11) gives us the following target potential

$$V(x) = -\left(\mathcal{E}_n/\sqrt{2}\tau\sigma\lambda^2\right)\mu(x)^{-1} + \frac{1}{8m}\left[\frac{m''}{m} - \frac{7}{4}\left(\frac{m'}{m}\right)^2 - \frac{3}{4\tau^2}\frac{m}{\mu^2}\right] \qquad (2.15)$$

Since the potential should be independent of the index $n$, we conclude from Eq. (2.15) that the parameter $\sigma$ is proportional to $\mathcal{E}_n$. From dimensional arguments and using the available parameters in the problem, we choose to write it as $\sigma = \left(\sqrt{2}\tau/\lambda^2\right)\mathcal{E}_n$. Thus, we end up with the following expression for the target potential

$$V(x) = -\frac{1}{2\tau^2}\mu(x)^{-1} + \frac{1}{8m}\left[\frac{m''}{m} - \frac{7}{4}\left(\frac{m'}{m}\right)^2 - \frac{3}{4\tau^2}\frac{m}{\mu^2}\right] \qquad (2.16)$$

Moreover, the transformation function now reads $q(x) = 2\lambda^{-1}\sqrt{\mu(x)/(2n+1)}$. The energy spectrum and eigenstates wavefunctions become

$$\begin{aligned}E_n &= -2/\tau^2(2n+1)^2 \\ \phi_n(x) &= A_n\left[m(x)\mu(x)\right]^{1/4} e^{-2\mu(x)/(2n+1)} H_n\left(2\sqrt{\mu(x)/(2n+1)}\right)\end{aligned} \qquad (2.17)$$

Note that $\sigma$, as expected, disappears from the final results.

## III. COULOMB AND MORSE POTENTIAL CLASSES

In this section we repeat briefly the same development, which was carried out above, for the Coulomb and Morse reference potentials. We start with the one-dimensional Coulomb problem, which has the following potential function, energy spectrum and wavefunctions [20]:

$$\mathcal{V}(y) = -Z/y$$
$$\mathcal{E}_n = -Z^2/2(n+1)^2 \qquad ; n = 0,1,2,\ldots \qquad (3.1)$$
$$\psi_n(y) = a_n y e^{-Zy/(n+1)} L_n^1\left(2Zy/(n+1)\right)$$

where $Z$ is the particle's charge number and $L_n^\nu(x)$ is the generalized Laguerre polynomial [21]. The two possible solutions are:

(Coulomb-1): $(q')^2 = m$
In this case we obtain, using Eq. (2.10), the following target potential

$$V(x) = -\alpha^2 \mu(x)^{-1} + \frac{1}{8m}\left[\frac{m''}{m} - \frac{7}{4}\left(\frac{m'}{m}\right)^2\right] \qquad (3.2)$$

where $\alpha$ is a real coupling parameter. The energy spectrum and wavefunctions are:

$$E_n = -(\alpha^2\tau)^2/(n+1)^2$$
$$\phi_n(x) = A_n m(x)^{1/4} \mu(x) e^{-\alpha^2\tau^2\mu(x)/(n+1)} L_n^1\left(2\alpha^2\tau^2\mu(x)/(n+1)\right) \qquad ; n = 0,1,2\ldots \qquad (3.3)$$

(Coulomb-2): $(q')^2 \mathcal{V}(q) = -m/\sigma^2$
In this case $\mathcal{W}(y) = 2\sqrt{Zy}$ giving $q(x) = (\tau/2\sigma)^2 Z^{-1} \mu(x)^2$. Substitution in Eq. (2.11) gives the following target potential

$$V(x) = \mathcal{E}_n\left(\tau/2Z\sigma^2\right)^2 \mu(x)^2 + \frac{1}{8m}\left[\frac{m''}{m} - \frac{7}{4}\left(\frac{m'}{m}\right)^2 - \frac{3}{\tau^2}\frac{m}{\mu^2}\right] \qquad (3.4)$$

Thus, we conclude that $\sigma^4$ is proportional to $\mathcal{E}_n$ and choose to write it as $\sigma^4 = -\frac{1}{2}\left(\tau^2/Z\right)^2 \mathcal{E}_n$ giving the transformation function $q(x) = \left[(n+1)/Z\right]\mu(x)^2$ and the target potential

$$V(x) = \frac{-1}{2\tau^2}\mu(x)^2 + \frac{1}{8m}\left[\frac{m''}{m} - \frac{7}{4}\left(\frac{m'}{m}\right)^2 - \frac{3}{\tau^2}\frac{m}{\mu^2}\right] \qquad (3.5)$$

The energy spectrum and wavefunctions, obtained using Eq. (2.11), are as follows:

$$E_n = -2(n+1)/\tau^2$$
$$\phi_n(x) = A_n m(x)^{1/4} \mu(x)^{3/2} e^{-\mu(x)^2} L_n^1\left(2\mu(x)^2\right) \qquad ; n = 0,1,2\ldots \qquad (3.6)$$

Next, we will address the Morse class of potentials. This class has the following reference potential, energy spectrum and wavefunctions [22]:

$$\mathcal{V}(y) = -\frac{\lambda^2}{2}\left(e^{-2\lambda y} - \xi\right)^2$$

$$\mathcal{E}_n = -\frac{\lambda^2}{2}(\xi - 2n - 1)^2 - \frac{\lambda^2 \xi^2}{2} \quad ; \quad n = 0,1,2,\ldots,n_{\max} \quad (3.7)$$

$$\psi_n(y) = a_n e^{-(\xi - 2n - 1)\lambda y} \exp\left(-\tfrac{1}{2} e^{-2\lambda y}\right) L_n^{\xi - 2n - 1}\left(e^{-2\lambda y}\right)$$

where $\lambda$ and $\xi$ are the Morse oscillator positive parameters and $n_{\max}$ is the largest integer which is $\leq \xi/2$. Due to the complicated expression of the transformation function in the case where $q(x) = \mathcal{W}^{-1}(\tau\mu(x)/\sigma)$, we will only consider the case where $q(x) = \tau\mu(x)$. For this latter case we obtain, using Eq. (2.10), the target problem described by the following:

$$V(x) = -\frac{\lambda^2}{2}\left[e^{-2\lambda\tau\mu(x)} - \xi\right]^2 + \frac{1}{8m}\left[\frac{m''}{m} - \frac{7}{4}\left(\frac{m'}{m}\right)^2\right]$$

$$E_n = \frac{\lambda^2}{2}(\xi - 2n - 1)^2 - \frac{\lambda^2 \xi^2}{2} \quad ; n = 0,1,2,\ldots,n_{\max} \quad (3.8)$$

$$\phi_n(x) = A_n m(x)^{1/4} e^{-(\xi - 2n - 1)\lambda\tau\mu(x)} e^{-f(x)/2} L_n^{\xi - 2n - 1}(f(x))$$

where $f(x) = e^{-2\lambda\tau\mu(x)}$.

For the sake of completeness, the three dimensional problem will be formally addressed. An illustrative 3D example for mass functions of powers of the radius will be given in the Appendix. The Hamiltonian (2.1) for a three dimensional problem with spherical symmetry gives the following radial wave equation:

$$\left\{\frac{d^2}{dr^2} - \frac{\ell(\ell + 1)}{r^2} + \frac{m'}{m}\left(\frac{1}{r} - \frac{d}{dr}\right) - 2m[V(r) - E]\right\}\phi(r) = 0 \quad (3.9)$$

where $\ell$ is the angular momentum quantum number, $m' \equiv dm(r)/dr$ and, as usual, we wrote the radial wavefunctions $\chi(r) = \phi(r)/r$. On the other hand, the time independent radial Schrödinger wave equation for a constant mass and angular momentum $\mathcal{L}$ reads

$$\left\{\frac{d^2}{d\rho^2} - \frac{\mathcal{L}(\mathcal{L} + 1)}{\rho^2} - 2[\mathcal{V}(\rho) - \mathcal{E}]\right\}\psi(\rho) = 0 \quad (3.10)$$

Acting on this last equation by the transformation $\rho = q(r)$ and $\psi(\rho) = g(r)\phi(r)$ results in the following wave equation

$$\left\{\frac{d^2}{dr^2} + \left(2\frac{g'}{g} - \frac{q''}{q'}\right)\frac{d}{dr} + \left(\frac{g''}{g} - \frac{q''}{q'}\frac{g'}{g}\right) - \mathcal{L}(\mathcal{L} + 1)\left(\frac{q'}{q}\right)^2 - 2(q')^2[\mathcal{V}(q(r)) - \mathcal{E}]\right\}\phi(r) = 0$$

Identifying this with the wave equation (3.9) gives the following

$$g(r) = \sqrt{q'/m} \quad (3.11)$$

$$V(r) - E + \frac{\ell(\ell + 1)}{2mr^2} = \frac{(q')^2}{m}[\mathcal{V}(q) - \mathcal{E}] + \frac{\mathcal{L}(\mathcal{L} + 1)}{2m}\left(\frac{q'}{q}\right)^2 + \frac{m'}{2m^2 r} + \frac{1}{m}F(m,q) \quad (3.12)$$

Therefore, given a radial effective mass function $m(r)$, Eq. (3.12) will be used to determine $V(r)$ and $E$ for a selected transformation function $q(r)$. This will be illustrated in one of the examples in the Appendix.

# APPENDIX: ILLUSTRATIVE EXAMPLES

We illustrate the utility of the PCT method being developed in this paper by few examples. For the sake of comparison with other methods some examples will be used to reproduce earlier results by other authors. However, new examples will also be presented to demonstrate the wide range and ease of application of the method.

**Example (1):** $m(x) = \left[(\gamma + x^2)/(1+x^2)\right]^2$, where $\gamma$ is a real constant parameter. This position dependent mass was studied in references [11] and [14]. We will give four different exactly solvable systems with this mass distribution. The potential function, energy spectrum and wavefunctions will be written down for all these systems. The integral in Eq. (2.9) gives $\mu(x) = \tau^{-1}\left[x + (\gamma - 1)\tan^{-1} x\right]$. Using the results of the development in sections II and III, we obtain the following:

(Oscillator-1):

$$V(x) = \frac{1}{2}\alpha^2 \mu(x)^2 + \left(\frac{\gamma - 1}{2}\right)\frac{3x^4 + 2(2-\gamma)x^2 - \gamma}{(\gamma + x^2)^4}$$

$$E_n = (\alpha/\tau)(n + \tfrac{1}{2})$$

$$\phi_n(x) = A_n \sqrt{\frac{\gamma + x^2}{1 + x^2}} e^{-\alpha\tau\mu(x)^2/2} H_n\left(\sqrt{\alpha\tau}\,\mu(x)\right)$$

(Oscillator-2):

$$V(x) = -\frac{1}{2\tau^2}\mu(x)^{-1} - \frac{3}{32\tau^2}\mu(x)^{-2} + \left(\frac{\gamma - 1}{2}\right)\frac{3x^4 + 2(2-\gamma)x^2 - \gamma}{(\gamma + x^2)^4}$$

$$E_n = -2/\tau^2(2n+1)^2$$

$$\phi_n(x) = A_n \sqrt{\frac{\gamma + x^2}{1 + x^2}} \mu(x)^{1/4} e^{-2\mu(x)/(2n+1)} H_n\left(2\sqrt{\mu(x)/(2n+1)}\right)$$

(Coulomb-1):

$$V(x) = -\alpha^2 \mu(x)^{-1} + \left(\frac{\gamma - 1}{2}\right)\frac{3x^4 + 2(2-\gamma)x^2 - \gamma}{(\gamma + x^2)^4}$$

$$E_n = -(\alpha^2 \tau)^2 / (n+1)^2$$

$$\phi_n(x) = A_n \sqrt{\frac{\gamma + x^2}{1 + x^2}} \mu(x) e^{-\alpha^2 \tau^2 \mu(x)/(n+1)} L_n^1\left(2\alpha^2 \tau^2 \mu(x)/(n+1)\right)$$

(Coulomb-2):

$$V(x) = \frac{-1}{2\tau^2}\mu(x)^2 - \frac{3}{8\tau^2}\mu(x)^{-2} + \left(\frac{\gamma - 1}{2}\right)\frac{3x^4 + 2(2-\gamma)x^2 - \gamma}{(\gamma + x^2)^4}$$

$$E_n = -2(n+1)/\tau^2$$

$$\phi_n(x) = A_n \sqrt{\frac{\gamma + x^2}{1 + x^2}} \mu(x)^{3/2} e^{-\mu(x)^2} L_n^1\left(2\mu(x)^2\right)$$

**Example (2):** We consider the smooth mass step $m(x) = 1 + \tanh(\gamma x)$, which becomes abrupt as $\gamma$ becomes large. This example was considered by L. Dekar *et al* [10] for a potential that has the same shape of a smooth step. Here, we are interested in the bound states. Therefore, we will select two exactly solvable systems with this mass step but for potentials that differ from that in reference [10]. In this case, Eq. (2.9) gives

$$\mu(x) = \left(\sqrt{2}/\tau\gamma\right)\tanh^{-1}\left(\sqrt{1+\tanh(\gamma x)}/\sqrt{2}\right)$$

The PCT method results in several solvable systems of which we choose the following:

(Oscillator-1):

$$V(x) = \frac{1}{2}\alpha^2\mu(x)^2 - \frac{7}{32\gamma^2}\frac{1+\tanh(\gamma x)}{\left[\sinh(\gamma x)+\cosh(\gamma x)\right]^2}$$

$$E_n = (\alpha/\tau)(n+\tfrac{1}{2})$$

$$\phi_n(x) = A_n\left[1+\tanh(\gamma x)\right]^{1/4} e^{-\alpha\tau\mu(x)^2/2} H_n\left(\sqrt{\alpha\tau}\mu(x)\right)$$

(Coulomb-1):

$$V(x) = -\alpha^2\mu(x)^{-1} - \frac{7}{32\gamma^2}\frac{1+\tanh(\gamma x)}{\left[\sinh(\gamma x)+\cosh(\gamma x)\right]^2}$$

$$E_n = -(\alpha^2\tau)^2/(n+1)^2$$

$$\phi_n(x) = A_n\left[1+\tanh(\gamma x)\right]^{1/4} \mu(x) e^{-\alpha^2\tau^2\mu(x)/(n+1)} L_n^1\left(2\alpha^2\tau^2\mu(x)/(n+1)\right)$$

**Example (3):** We consider the asymptotically vanishing mass $m(x) = (\gamma + x^2)^{-1}$. In this case $\mu(x) = \tau^{-1}\ln\left(x+\sqrt{\gamma+x^2}\right)$. Among the many possible exactly solvable systems with this mass that could be obtained by the PCT method, we may choose the following:

(Oscillator-1):

$$V(x) = \frac{1}{2}\alpha^2\mu(x)^2 - \frac{1}{8}\frac{x^2+2\gamma}{x^2+\gamma}$$

$$E_n = (\alpha/\tau)(n+\tfrac{1}{2})$$

$$\phi_n(x) = A_n(\gamma+x^2)^{-1/4} e^{-\alpha\tau\mu(x)^2/2} H_n\left(\sqrt{\alpha\tau}\mu(x)\right)$$

**Example (4):** We consider in this example the Morse class and take $m(x) = \tanh(\lambda x)^2$. This gives $\mu(x) = (\lambda\tau)^{-1}\ln(\cosh(x))$. Using the set of equations (3.8) we obtain the target system defined by the following set of quantities:

$$V(x) = -\frac{\lambda^2}{2}\left[\cosh(\lambda x)^{-2} - \xi\right]^2 - \frac{\lambda^2}{2}\left[\sinh(\lambda x)^{-2} + (5/4)\sinh(\lambda x)^{-4}\right]$$

$$E_n = \frac{\lambda^2}{2}(\xi - 2n - 1)^2 - \frac{\lambda^2\xi^2}{2}$$

$$\phi_n(x) = A_n\sqrt{|\tanh(\lambda x)|}\left[\cosh(\lambda x)\right]^{-(\xi-2n-1)} e^{-1/2\cosh(\lambda x)^2} L_n^{\xi-2n-1}\left(\cosh(\lambda x)^{-2}\right)$$

**Example (5):** In this last example we tackle the three dimensional problem and take the 3D isotropic oscillator as reference, which is defined by the following [20]:

$$\mathcal{V}(\rho) = \tfrac{1}{2}\lambda^4 \rho^2$$
$$\mathcal{E}_n = \lambda^2(2n + \mathcal{L} + \tfrac{3}{4})$$
$$\psi_n(\rho) = a_n (\lambda\rho)^{\mathcal{L}+1} e^{-\lambda^2 \rho^2/2} L_n^{\mathcal{L}+\frac{1}{2}}\left(\lambda^2 \rho^2\right)$$

We consider the radial dependent mass $m(r) = \alpha r^\gamma$ and take the PCT function $q(r) = r^\nu$, where $\alpha$, $\gamma$ and $\nu$ are nonzero real parameters. By substitution in Eq. (3.12) we obtain the following two solutions:

(a) $\nu = 1 + \gamma/2$ and $\gamma \neq -2$:
$$V(r) = \frac{\alpha}{2} C^2 r^{\gamma+2}$$
$$E_n = \frac{\gamma+2}{2} C(2n + \Lambda + 1/4)$$
$$\phi_n(r) = A_n (\xi r)^{(1+\gamma/2)\Lambda + (\gamma+1)/2} \exp\left(-(\xi r)^{\gamma+2}/2\right) L_n^\Lambda \left((\xi r)^{\gamma+2}\right)$$

where $\Lambda(\ell) = |\gamma+2|^{-1} \sqrt{4\ell(\ell+1) + (\gamma-1)^2}$, $C$ is a real potential coupling parameter and 
$$\xi = \left(\frac{2\alpha C}{\gamma+2}\right)^{\frac{1}{\gamma+2}}.$$

(b) $\nu = 1/2 + \gamma/4$ and $\gamma \neq -2$:
$$V(r) = -\frac{1}{2}\frac{C}{\sqrt{\alpha}} r^{-1-\gamma/2}$$
$$E_n = -\frac{C^2/2}{(\gamma+2)^2} \frac{1}{(n+\Lambda+1/8)^2}$$
$$\phi_n(r) = A_n (\xi_n r)^{(1+\gamma/2)\Lambda + (1+\gamma)/2} \exp\left(-(\xi_n r)^{1+\gamma/2}/2\right) L_n^{2\Lambda}\left((\xi_n r)^{1+\gamma/2}\right)$$

where $\xi_n = \left[\dfrac{4\sqrt{\alpha} C}{(\gamma+2)^2 (n+\Lambda+1/8)}\right]^{\frac{1}{1+\gamma/2}}$ and $\Lambda(\ell)$ as above.

The singular case $m(r) = \alpha r^{-2}$, which corresponds to $\gamma = -2$, is solved by taking $q(r) = \lambda^{-1} \ln(r)$ giving the following S-wave ($\ell = 0$) solution:
$$V(r) = \frac{1}{2\alpha} \ln(r)^2 + \frac{C^2}{2} \ln(r)^{-2}$$
$$E_n = \alpha^{-1}(2n + \Lambda + 11/8)$$
$$\phi_n(r) = A_n r^{-1/2} \ln(r)^{\Lambda + 1/2} \exp\left(-\ln(r)^2/2\right) L_n^\Lambda \left(\ln(r)^2\right)$$
where $\Lambda = 2^{-1}\sqrt{1 + 4\alpha C^2}$.

**REFERENCES:**

[1] See, for example, G. A. Natanzon, Teor. Mat. Fiz. **38**, 146 (1979); L. E. Gendenshtein, Zh. Eksp. Teor. Fiz. Pis'ma Red. **38**, 299 (1983) [JETP Lett. **38**, 356 (1983)]; F. Cooper, J. N. Ginocchi, and A. Khare, Phys. Rev. D **36**, 2438 (1987); R. Dutt, A. Khare, and U. P. Sukhatme, Am. J Phys. **56**, 163 (1988); *ibid.* **59**, 723 (1991); G. Lévai, J Phys. A **22**, 689 (1989); *ibid.* **27**, 3809 (1994); R. De, R. Dutt, and U. Sukhatme, J Phys. A **25**, L843 (1992)

[2] See, for example, M. F. Manning, Phys. Rev. **48**, 161 (1935); A. Bhattacharjie and E. C. G. Sudarshan, Nuovo Cimento **25**, 864 (1962); N. K. Pak and I. Sökmen, Phys. Lett. **103A**, 298 (1984); H. G. Goldstein, *Classical Mechanics* (Addison-Wesley, Reading-MA, 1986); R. Montemayer, Phys. Rev. A **36**, 1562 (1987); G. Junker, J Phys. A **23**, L881 (1990)

[3] See, for example, E. Witten, Nucl. Phys. B **185**, 513 (1981); F. Cooper and B. Freedman, Ann. Phys. (NY) **146**, 262 (1983); C. V. Sukumar, J. Phys. A **18**, 2917 (1985); A. Arai, J. Math. Phys. **30**, 1164 (1989); F. Cooper, A. Khare, and U. Sukhatme, Phys. Rep. **251**, 267 (1995)

[4] See, for example, B. G. Wybourne, *Classical groups for physicists* (Wiley-Interscience, New York, 1974), and references therein; W. Miller Jr., *Lie theory and special functions* (Academic, New York, 1968); Y. Alhassid, F. Gürsey, and F. Iachello, Phys. Rev. Lett. **50**, 873 (1983); Ann. Phys. (N.Y.) **148**, 346 (1983) ; *ibid.* **16**7, 181 (1986); Y. Alhassid, F. Iachello, and R. Levine, Phys. Rev. Lett. **54**, 1746 (1985); Y. Alhassid, F. Iachello, and J. Wu, Phys. Rev. Lett. **56**, 271 (1986); J. Wu and Y. Alhassid, J. Math. Phys. **31**, 557 (1990); M. J. Englefield and C. Quesne, J. Phys. A **24**, 3557 (1991)

[5] See, for example, A. de Souza-Dutra, Phys. Rev. A **47**, R2435 (1993); N. Nag, R. Roychoudhury, and Y. P. Varshni, Phys. Rev. A **49**, 5098 (1994); R. Dutt, A. Khare, and Y. P. Varshni, J. Phys. A **28**, L107 (1995); C. Grosche, J. Phys. A, **28**, 5889 (1995); *ibid.* **29**, 365 (1996); G. Lévai and P. Roy, Phys. Lett. A **270**, 155 (1998); G. Junker and P. Roy, Ann. Phys. (N.Y.) **264**, 117 (1999); R. Roychoudhury, P. Roy, M. Zonjil, and G. Lévai, J. Math. Phys. **42**, 1996 (2001)

[6] See, for example, A. V. Turbiner, Commun. Math. Phys. **118**, 467 (1988); M. A. Shifman, Int. J. Mod. Phys. A **4**, 2897 (1989); R. Adhikari, R. Dutt, and Y. Varshni, Phys. Lett. A **141**, 1 (1989); J. Math. Phys. **32**, 447 (1991); R. K. Roychoudhury, Y. P. Varshni, and M. Sengupta, Phys. Rev. A **42**, 184 (1990); L. D. Salem and R. Montemayor, Phys. Rev. A **43**, 1169 (1991); M. W. Lucht and P. D. Jarvis, Phys. Rev. A **47**, 817 (1993); A. G. Ushveridze, *Quasi-exactly Solvable Models in Quantum Mechanics* (IOP, Bristol, 1994)

[7] A. D. Alhaidari, Phys. Rev. Lett. **87** , 210405 (2001); **88**, 189901 (2002)

[8] A. D. Alhaidari, J. Phys. A **34**, 9827 (2001); **35**, 3143 (2002)

[9] A. D. Alhaidari, Phys. Rev. A **65**, 042109 (2002); **66**, 019901 (2002)

[10] L. Dekar, L. Chetouani and T. F. Hammann, J. Math. Phys. **39**, 2551 (1998); Phys. Rev. A **59**, 107 (1999)

[11] A. R. Plastino, A. Rigo, M. Casas, F. Gracias and A. Plastino, Phys. Rev. A **60**, 4318 (1999)

[12] V. Milanović and Z. Iković, J. Phys. A **32**,7001 (1999)

[13] A. de Souza Dutra and C. A. S. Almeida, Phys. Lett. A **275**, 25 (2000)

[14] B. Roy and P. Roy, J. Phys. A **35**, 3961 (2002)

[15] G. Bastard, *Wave Mechanics Applied to Semiconductor Heterostructure* (Les Editions de Physique, Les Ulis, France, 1988)

[16]  L. Serra and E. Lipparini, Europhys. Lett. **40**, 667 (1997)

[17]  M. Barranco, M. Pi, S. M. Gatica, E. S. Hernandez, and J. Navarro, Phys. Rev. B **56**, 8997 (1997)

[18]  F. Arias de Saavedra, J. Boronat, A. Polls, and A. Fabrocini, Phys. Rev. B **50**, 4248 (1994)

[19]  See, for example, T. Gora and F. Williams, Phys. Rev. **177**, 1179–1182 (1969); O. Von Roos, Phys. Rev. B **27**, 7547–7552 (1983); O. Von Roos and H. Mavromatis, Phys. Rev. B **31**, 2294–2298 (1985); R. A. Morrow, Phys. Rev. B **35**, 8074–8079 (1987); **36**, 4836–4840 (1987); W. Trzeciakowski, Phys. Rev. B **38**, 4322–4325 (1988); I. Galbraith and G. Duggan, Phys. Rev. B **38**, 10057–10059 (1988); K. Young, Phys. Rev. B **39**, 13434–13441 (1989); G. T. Einevoll, P. C. Hemmer, and J. Thomsen, Phys. Rev. B **42**, 3485–3496 (1990); G. T. Einvoll, Phys. Rev. B **42**, 3497–3502 (1990); C. Weisbuch, B. Vinter, *Quantum Semiconductor Heterostructures* (Academic Press, New York, 1993)

[20]  See any standard book in quantum mechanics, for example, A. Messiah, *Quantum Mechanics I & II* (Wiley, New York, 1966); E. Merzbacher, *Quantum Mechanics*, 2nd ed. (Wiley, New York, 1970)

[21]  W. Magnus, F. Oberhettinger, and R. P. Soni, *Formulas and Theorems for the Special Functions of Mathematical Physics*, 3rd edition (Springer-Verlag, New York, 1966)

[22]  P. M. Morse, Phys. Rev. **34**, 57 (1929)